# Effective Metagraph-based Life Pattern Clustering with Big Human Mobility Data

Wenjing Li[1,2], Haoran Zhang[2,1*], Jinyu Chen[1], Peiran Li[1], Yuhao Yao[1], Mariko Shibasaki[1], Xuan Song[1,3], Ryosuke Shibasaki[1]

[1]Center for Spatial Information Science, The University of Tokyo, 5-1-5 Kashiwanoha, Kashiwa-shi, Chiba, Japan, 277-8568

[2]LocationMind Inc., 701, 3-5-2, Iwamoto-cho, Chiyoda-ku, Tokyo, Japan, 101-0032

[3]Southern University of Science and Technology-University of Tokyo Joint Research Center for Super Smart Cities, Department of Computer and Engineering, Southern University of Science and Technology, 518055 Shenzhen, Guangdong, China

*Correspondence: zhang_ronan@csis.u-tokyo.ac.jp

**Abstract:**

Life pattern clustering is essential for abstracting the groups' characteristics of daily mobility patterns and activity regularity. Based on millions of GPS records, this paper proposed a framework on the life pattern clustering which can efficiently identify the groups have similar life pattern. The proposed method can retain original features of individual life pattern data without aggregation. Metagraph-based data structure is proposed for presenting the diverse life pattern. Spatial-temporal similarity includes significant places semantics, time sequential properties and frequency are integrated into this data structure, which captures the uncertainty of an individual and the diversities between individuals. Non-negative-factorization-based method was utilized for reducing the dimension. The results show that our proposed method can effectively identify the groups have similar life pattern and takes advantages in computation efficiency and robustness comparing with the traditional method. We revealed the representative life pattern groups and analyzed the group characteristics of human life patterns during different periods and different regions. We believe our work will help in future infrastructure planning, services improvement and policies making related to urban and transportation, thus promoting a humanized and sustainable city.

## 1.Introduction

The concerned field of Urban Computing integrates computational solutions to study how humans travel and conduct daily activities, making cities more humanized, more livable, and more sustainable (Song, et al., 2020; Y. Zheng, Capra, Wolfson, & Yang, 2014). Life pattern, which describes the individual's activity regularity and general travel style, has gained significant attention by city and transportation planners and urban policymakers. However, individual daily activities are very diverse(Fu, McKenzie, Frias-Martinez, & Stewart, 2018). Abstracting the groups' characteristics of life patterns from diverse individual cases is essential for urban infrastructure improvement and policies making towards multiple types of citizens. Life pattern clustering is the core of abstracting the groups' characteristics of mobility patterns.

Clustering life pattern is challenging. When regarding the life pattern of one person in one day, the life pattern data is high-dimensional. It includes temporal and spatial information, and this two-dimension information is coupled. When considering the life pattern of one person in multiple periods, the individual life pattern records contain massive uncertainties. One person can make different options for daily activity types and travel modes at different times and periods. When regarding the life pattern of multiple people in multiple periods, there are considerable diversities in life patterns. Life pattern varies between differences groups, region and time. The condition and situation will be more complex and diverse if regarding in a larger sampling scale. These put difficulties in life pattern clustering. The universal life pattern clustering methods usually aggregate the features from individual life patterns, however, much information is lost during this aggregation process.

In this research, based on millions of GPS records, we proposed a framework on the life pattern clustering which can efficiently identify the groups have similar life pattern. The proposed metagraph-based method can lossless retain original features of individual life pattern data without aggregated and handle the high-dimensional, uncertain and diverse life pattern data. Spatial-temporal similarity includes significant places semantics, time sequential properties and frequency are considered in the proposed life pattern method. We believe our work help in future infrastructure planning, services improvement and policies making related to urban and transportation, thus promoting a humanized and sustainable city.

The rest of the paper is organized as follows. In section 2, we review the previous works on life pattern clustering. The detailed description of the problem we focus on is presented in Section 3. In Section 4, we introduce the proposed method to cluster life patterns from raw GPS data. Section 4 introduces the study case and the dataset we utilize. The result and comprehensive analysis are presented in Section 6. Finally, conclusions and future work are given in Section 7.

## 2. Related works

Life pattern is one of the most popular topics in the field of urban computing, urban planning, transportation, and so on. Some studies may refer it in terms of "mobility pattern" (Jiang, Ferreira, & Gonzalez, 2017; K. Liu, Murayama, & Ichinose, 2021; Tu, et al., 2020), "activity pattern"(Hafezi, Liu, & Millward, 2019; Phithakkitnukoon, Horanont, Di Lorenzo, Shibasaki, & Ratti, 2010; W. Zhang & Thill, 2017) or "commute pattern"(Acheampong, 2020; Yu, Li, Yang, & Zhang, 2020). But they all discuss how humans travel and conduct daily activities. The routine mobility between home and workplaces are commonly discussed, other places include tourist attraction (Phithakkitnukoon, et al., 2015), shopping (Hu, Song, Wang, Xie, & Luo, 2016) and catering (F. Zhang, et al., 2016) also be subdivided into some studies.

In the past, life pattern-related researches were mainly based on the statistics results from official records or questionnaire surveys. Kahneman, Krueger, Schkade, Schwarz, and Stone (2004) conducted a Day Reconstruction Method by survey to study how people spend their time. Matz, et al. (2014) conducted a national survey to study the Canadian Human Activity Pattern and the effort factors. However, the relatively small sampling data would bring bias in representing all characteristics of the overall research population (J. Chen, et al., 2020; Yao, et al., 2020). They are not enough to reveal groups' characteristics of life patterns at a large scale. The similarity and differences of life patterns are also remained to be furtherly quantified. Recently, with the development of information and communication technologies, big data with location provide a new horizon to life pattern for urban research and enable us to analyze large-scale life patterns from the real-world. With smart card data from Santiago in Chile and Gatineau in Canada, Devillaine, Munizaga, and Trépanier (2012) detected the activity purpose and time of public transport trips.Poucin, Farooq, and Patterson (2018) utilized Wi-Fi data to mine the human main activates in large area. Jiang, et al. (2017) utilized CDR (call detail record) data to capture individual activity-based behavior and mobility patterns. Qin, Shangguan, Song, and Tang (2018) developed a Spatio-Temporal Routine Mining Model to describe the human routine behaviors from mobile phone data. Among all kinds of big data with location information, GPS (Global Position System ) data is considered to have a high frequency and accuracy in recording the individual location history (V. W. Zheng, Zheng, Xie, & Yang, 2010). Ye, Zheng, Chen, Feng, and Xie (2009) proposed a framework to systematically describe how to mine life regularity from GPS data. Their work focused on significant places of individual life and integrated diverse properties into significant places. X. Chen, Shi, Zhao, and Liu (2016) mined people's periodic activity patterns from massive GPS data. Li, Hu, Dai, Fan, and Wu (2020) gave a portrait depiction of individual mobility patterns using mobile phone GPS data. Long and Reuschke (2021) utilized GPS smartphone to examine the daily mobility patterns of small business owners and homeworkers.

An important aspect of life patterns is life pattern clustering. From the aspect of similarity metric, most of the research utilized statistical features mining from trip data. With a fixed number of clusters, Ordóñez Medina and Erath (2013) categorize workers and their daily work activities according to start time and duration. El Mahrsi, Côme, Baro, and Oukhellou (2014) recognized weekly traveling patterns of public transport users and clustered passengers based on temporal profiles. Ma, Wu, Wang, Chen, and Liu (2013) represented the regularity of users by 4 features: number of travel days, number of similar first boarding times, number of similar route sequences and number of similar stop ID sequences. K-means ++ algorithm was applied to find clusters of similar users. Medina (2018) utilized public transport smart card data and household travel surveys to infer the weekly primary activity patterns of individuals in Singapore. She used 14 features to summarize the users' primary activity pattern and applied DBSCAN algorithm for user clustering. As the number of measure metrics is preset and limited, they are not enough to reflect the temporal modes of life pattern between different groups. Besides, they cannot describe the coupled relationship between space and temporality. Some cluster life patterns which share the similar location with high frequency. J.-G. Lee, Han, and Whang (2007) proposed a clustering algorithm that grouped similar trajectories as a whole. Yang, Cheng, and Chen (2018) utilized GPS data to mine the individual similarity with common significant places. This work emphasized the spatial-temporal similarity between individuals and their personally significant places. However, the sequential properties are ignored and they could not measure the temporal contextual similarities. Y. Zheng, Zhang, Ma, Xie, and Ma (2011) proposed a clustering method considering the sequence of movements, the hierarchical properties of the geographic space, and the popularity of the visited places. Based on 3 GPS datasets, Xu, et al. (2018) proposed a method which captures the semantic feature to detect the popular temporal modes. This method considers the contextual similarities and enables the individuals with similar temporal modes yet having large spatial distance to own similar representations. But some information such as frequency of travel at different time are missing during the aggregation process. **Table 1** gives an overview of similarity measurement of life pattern clustering in related literature studies.

**Table 1**. An overview of similarity measurement of life pattern clustering in related literature studies

| similarity measurement | references | features | dataset type | study area |
|---|---|---|---|---|
| Aggregate statistical features | Ordóñez Medina and Erath (2013) | start time and duration of work | Household Interview Transport Survey | Singapore |
| | El Mahrsi, et al. (2014) | the number of trips that the user took over each hour of each day of the week and the probability | smart card data | Rennes, France |
| | Ma, et al. (2013) | number of travel days, number of similar first boarding times, number of similar route sequences, number of similar stop ID sequences | smart cards | Beijing, China |
| | Medina (2018) | the start time and the duration of the activity of each day over a week | Household Interview Transport Survey and smart card data | Singapore |

| cluster life patterns that share similar location with high frequency | J.-G. Lee, et al. (2007) | graphic similarity of trajectory | trajectory data sets include timestamp, longitude and latitude | Atlantic |
|---|---|---|---|---|
| | Yang, et al. (2018) | share common significant places | GPS data | Beijing, China |
| the temporal contextual similarities | Y. Zheng, et al. (2011) | the hierarchical properties of the geographic space, the sequence of movements, the popularity of the visited places | GPS data | Beijing, China |
| | Xu, et al. (2018) | The temporal contextual modes | GPS data | Shanghai, China |

In the field of data sciences, some other researches focus on clustering. Han, Cheng, Xin, and Yan (2007) presented a comprehensive illustration of frequent pattern mining categories, techniques and applications. Lu and Wu (2015) and Lu and Miao (2016) proposed a tree-structured data parameterization method which can conduct high-dimension data clustering task on matrix manifold. It transforms tree structure data to a point in a low dimensional space. These works gave us the inspiration of how to handle high-dimension data clustering and represent multiple features such as spatial location, sequential properties and sequential properties.

Following this, the contributions of this work are shown as following:

(1) We proposed a novel life pattern clustering method that retains original features of individual life pattern data and considers the coupled spatial-temporal similarity includes significant places semantics, time sequential properties and frequency.
(2) Metagraph-based data structure and non-negative-factorization-based method are introduced to improve the efficiency and performances of the clustering of the high-dimensional life pattern data.
(3) The proposed method shows advantages in computation efficiency and robustness when handling the high-dimensional life pattern data.
(4) Based on millions of GPS trajectory records, we revealed the representative life pattern groups based on the clustering result and analyzed the spatial-temporal characteristics of life patterns in a large region.

## 3. Problem description

Despite some degree of change and spontaneity, human mobility shows a high degree of temporal and spatial regularity (Gonzalez, Hidalgo, & Barabasi, 2008). This inherent regularity enables us to summary the reproducible patterns of an individual's life. There are some similarities in daily activity patterns within the group of people (Phithakkitnukoon, et al., 2010). For example, a typical life pattern of office employees on workdays is that go to work in the morning and return home in the evening. A housewife may usually stay home all day. Resembling the definitions of the life pattern in related references (Ye, et al., 2009), we defined life patterns as individual activity regularity and general travel style. Since location history data such as GPS has a close relationship between people's daily life and geographic locations, we can discover individual life patterns from one's location history. Clustering life patterns and mining the group characteristics of life patterns on a large-scale from unlabeled mobility data is challenging. In this paper, we need to overcome three major challenges for clustering individual life pattern:

**Cluster the life pattern data in a high dimension:** from the aspect of one person one day, life pattern information, which couples temporal and spatial information, is high-dimensional. As the number of points required to represent any distribution exponentially increases with the number of dimensions, the data sets in high-dimensional tend to be sparse. In such a high-dimensional space, clustering becomes difficult due to the increasing sparsity of data. Besides, distinguishing distances between data points in high-dimensional space poses difficulties (Tomasev, Radovanovic, Mladenic, & Ivanovic, 2013). The metric to properly measure the similarities between groups should be carefully introduced.

**Handling the uncertainty in the life pattern of an individual**: if regarding the life pattern of one person in multiple periods, due to the uncertainty of human movements, the individual life pattern records extracted from GPS data contain massive uncertainties. Even one person can make different options for daily activity types, travel modes, time and periods.

**Represent the diverse life pattern between individuals in a suitable format:** Achieving a suitable format for life pattern clustering is a challenging task. First, the individual daily life pattern is diverse. if regarding the life pattern of multiple persons in multiple periods, life pattern varies between differences individual, region and time. When regarding on a larger scale, the condition and situation will be more complex and uncertain. Thus, the information of time, activity semantics, temporal order and frequency should be reflected in the presentation format of life pattern. The original features of individual life pattern data should be retained as much as possible. Besides, the metric to measure the similarities of life patterns should be computable through the representations.

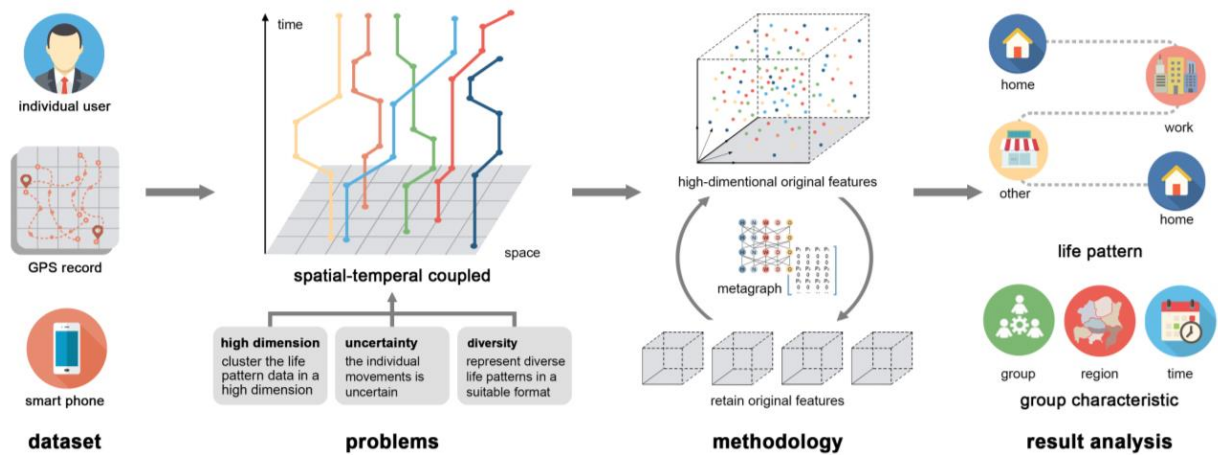

**Figure 1**. Framework of metagraph-based life pattern clustering

## 4.Methodology

The main goal of this paper is to provide a reliable method for life pattern clustering and mining the group characteristics of life patterns in large-scale unlabeled mobility data. To solve the above challenges respectively, we proposed a methodology that can effectively address the identified challenges. The methodology includes three processes: significant places detection, probability T-A matrix construction and matrix factorization. The whole process of the methodology is shown in **Figure 2**.

First, we conduct a three-step significant places detection from raw GPS records based on our previous work (H. Zhang, et al., 2018). Significant places are classified into 5 categories—home, work, daytime spot, nighttime spot and other places—for mining the individual activity regularity. Secondly, we proposed a metagraph-based method that can transfer the diverse individual life pattern into a unified form of the matrix. A support graph which covers all life pattern situation of each hour of all users is built. It captures the properties of time, activity semantics and temporal order. We transfer it into a probability T-A matrix and the information of frequency also be added in. Thirdly, concat all users' average probability T-A matrix and conduct non-negative matrix factorization to transfer the individual's life pattern into a point in a metagraph space. The dimension can be effectively reduced. Besides, once the geodesic distance is defined in the metagraph space, Euclidean distance can be introduced to measure the similarities between the points. Traditional clustering methods such as K-means can also be incorporated into the metagraph space. We will elaborate on each component of the methodology in the following section. Finally, we will give a comprehensive analysis of the life pattern clustering result from the aspect of group characteristics, region and time.

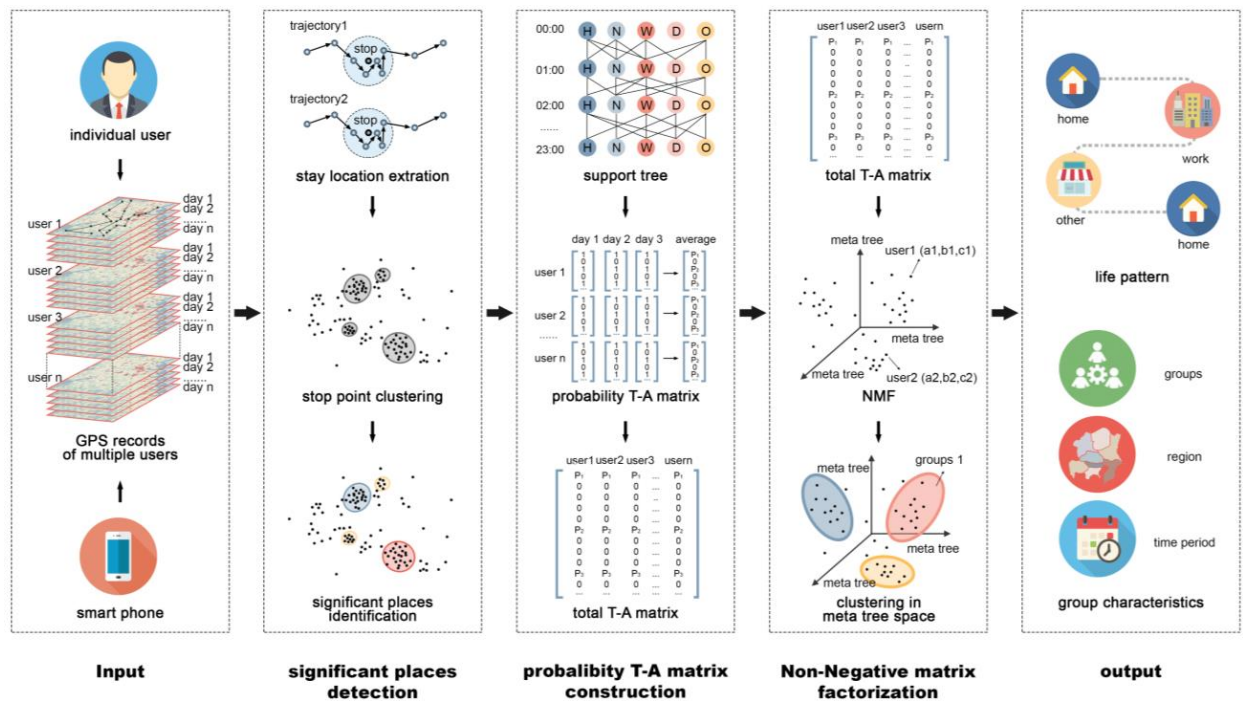

**Figure 2**. Methodology

## 4.1 Significant places detection

To identify life patterns from GPS trajectories, we conduct a significant place detection to mine the typical activity of individuals. The detection process includes three steps: stay location extraction, spatial clustering of stay location and significant places identification.

The first step was to extract *stay location* of the individual's raw human trajectory. The *stay location* of an individual's raw mobility trajectory is a series of consecutive points that represent the user staying at a location, while the moving location represents the user moving. The stay locations represent the places that a user spends a considerable amount of time, for example, home, workplace, restaurant, market, gym, park, etc. In this paper, the stay location extraction method we utilized is based on our previous work (H. Zhang, et al., 2018). It extracted the stay location according to the distance $\Delta l$ and time spam $\Delta t$ of temporal neighboring points which are less than the threshold. If $X_u = \{x_{t1}, x_{t2}, \ldots, x_{ti}, \ldots\}$ denotes a set of GPS locations of user u where $x_{ti}$ is the location at time ti, $\{x_{ti}, x_{ti+1}, x_{ti+2}, \ldots, x_{tm}\}$ that are within $\Delta l$ and $t_m - t_i \leqslant \Delta t$ are grouped as a stop (**Figure 3**). The ones which are greater than the threshold are identified as moving segments. The noise points were removed according to the mean and standard deviation of Gaussian distribution. By implementing this process, we determined the travel state of each user during each period and obtained the stay location which allows life pattern detection next step. Here, we used $S_u = \{S_1, S_2, S_3, \ldots, S_i\}$ to represent the user $u$'s stay location history. Each stay point $S_i$ represents a stay location. It retains the arrival time ($S.arvT$) and the leaving time ($S.levT$) which respectively equals to the timestamp of the first and last GPS point constructing this stay point.

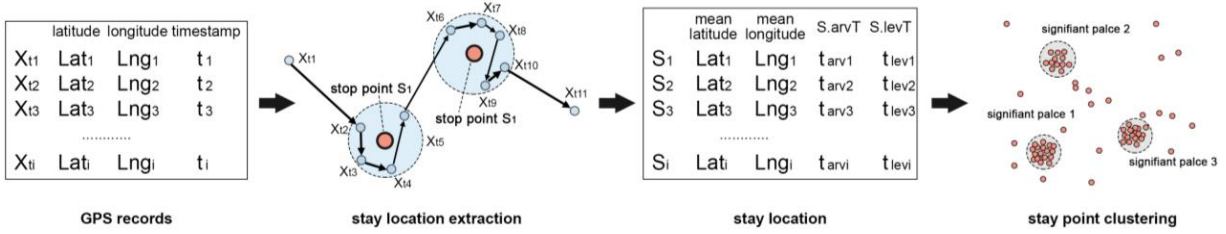

**Figure 3**. The illustration of the *stay location*

The second step was the spatial clustering of stay location. Due to the uncertainty of human movements, two stay points in the same places may have different spatial coordinates. Thus we need to conduct spatial clustering methods to group different stay points with the same semantic meaning. The centroid of the cluster was considered as a significant place (e.g., home, workplace, other)which represents the individual's typical activity (Ye, et al., 2009). Here, DBSCAN (Density-Based Spatial Clustering of Applications with Noise) was utilized for spatial clustering. DBSCAN has been adopted in many mining applications of spatial data. It is a highly popular clustering method in determining the zones regularly visited by the individuals (Comito, Falcone, & Talia, 2016; X. Liu, Huang, & Gao, 2019; Xie, 2019) and has been proved to have a good performance among other techniques (Phithakkitnukoon, et al., 2015). Two parameters are required in DBSCAN method: the minimum number of points that can form a cluster (*minpts*) and the maximum distance within which two points belong to a cluster (*eps*). Points with a neighborhood *eps* more than *minpts* are categorized as core points, and points reachable by a core point are border points of the cluster. Both core points and border points are considered as the clustered points. Points not within the *eps* search radius of any core point are treated as noise (Khan, Rehman, Aziz, Fong, & Sarasvady, 2014).

The last step was to identify significant places. Here, we assume that every person has his/her own home and most people return homes at the end of a day. Although there are exceptions in the real world, this assumption simplifies the way we identify the significant places from unlabeled GPS data and allows us to estimate life patterns justifiably. We classify the significant places into 5 groups — home(H), other frequently visit nighttime spots (N), workplaces (W), other frequently visit daytime spots (D), and other significant places (O). Within each cluster, we extracted the stay points which meet the following conditions as candidate points of nighttime spots or daytime spots : (1) candidates points of nighttime spots: between 8.pm to 6.am, the duration is larger than 1.5 h, more than 10 day's records within one cluster; (2) candidates points of daytime spots: between 9.am to 7.pm in the workday, the duration is larger than 1.5h, more than 10 day's record within one cluster. Then, we compared the day of stop and the total stop duration for daytime spot candidates and nighttime spot candidates within each cluster to determine the semantic meaning. The cluster whose stop days and stops duration of nighttime spot candidates are highest will be identified as home, and the cluster whose stop days and stops duration of daytime spot candidates are highest will be identified as workplaces. Other clusters that contain daytime spot candidates or nighttime spot candidates will be identified as N or D. The cluster without daytime spot candidates or nighttime spot candidates that meet the conditions will be identified as other significant places. After obtaining the estimation of home, workplace and other places, we can present the user *u*'s stay location at day *j* as $S_{u,j} \in S_u$, $S_{u,j} = \{S_{(1,j)}, S_{(2,j)}, home_{(k,j)}, S_{(3,j)}, \ldots, work_{(k,j)}, \ldots, Other_{(k,j)}, \ldots, S_{(i,j)}\}$, $Home_k \in H_u = \{home_1, home_2, home_3, \ldots, home_k\}$, $work_k \in W_u = \{work_1, work_2, work_3, \ldots, work_k\}$, $other_k \in O_u = \{other_1, other_2, other_3, \ldots, other_k\}$.

## 4.2 Support graph and Topology-attribute matrix construction

We illustrate how to present the life pattern of all users using a support graph and topology-attribute matrix (T-A matrix) in this section.

For each user, we first build an individual graph that represents his location of each hour within one day. The nodes of the individual graph assume to begin from the top and one layer of the individual graph represents one hour, each node of the graph represents one significant place for a user within one hour. For users which stay at multiple significant places within one hour, we select the one with the longest duration. There is only one edge between the same source and target nodes. The direction of the edge is always from the node in the higher layer to the node in the lower layer.

Secondly, a support graph is constructed in the way that any of the individual graphs can always be found in the support graph to have the same topology. In the support graph, each edge is assigned a unique index in an ascending manner from top to bottom and from left to right. Then, the edges of the individual graph will be respectively assigned one index which corresponds to the index of the edges with the same topology in the support graph by definition. **Figure 4 (a)** illustrates the construction of the individual graph, support graph and the edge indexing strategy based on a population of four individuals.

Then, we use T-A matrix to represent the graphs. Let $n$ denotes the number of edges in the support graph, a T-A matrix of one user one day will be generated with its row number being equal to the total index number of the support graph edges $n$. The first row of the T-A matrix corresponds to the first edge of the support graph, the second row to the second edge, and so forth. For individual each day, the element in the $i$-th row is assigned the value 1 if the ith edge of the support graph is also contained in the individual graph; otherwise, 0 is assigned to the element in the $i$-th row. Let $T_u = \{T_{(u,1)}, T_{(u,2)}, \ldots, T_{(u,m)}\}$ corresponding to T-A matrix of user $u$ in day $m$, by computing the average value of $T_u$, we can obtain the average probability matrix $T_{average\ u}$ of user $u$ with $m$ days. **Figure 4(b)** is an illustration of the T-A matrix generation based on user 2.

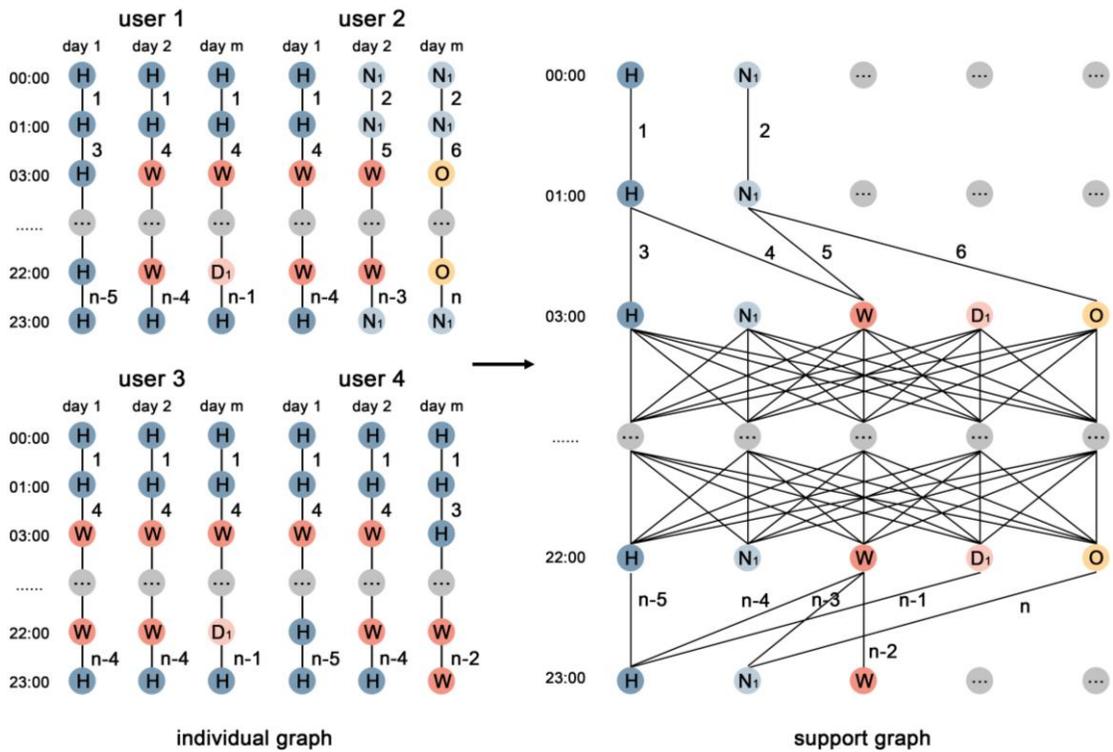

(a) Support graph construction and the branch indexing strategy based on a population of four individuals

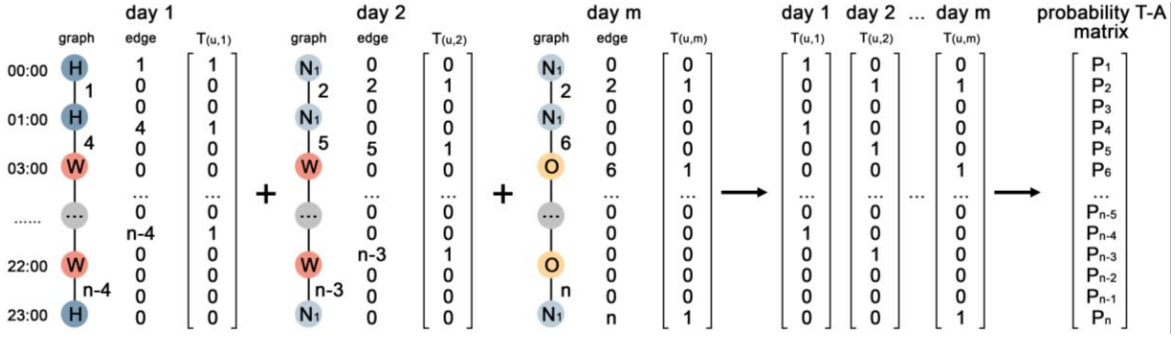

**(b)** T-A matrix generation based on user 2

**Figure 4**. Individual graph, support graph and Topology-attribute matrix construction

### 4.3 Structure constrained NMF and meta-graph space

For a population of average probability T-A matrix T = {$T_{average\ 1}$, $T_{average\ 2}$, . . . , $T_{average\ u}$}, *n* denote the support graph with a total of *n* edges, *u* denote the total number of users. A total T-A matrix T$n$×u can be constructed for the total graph. With this representation, we can conduct graph clustering in the matrix space M. Once we identify the bases of M, each $T_{average\ u}$ can be decomposed into a linear combination of these bases. Then, the similarity between a pair of graphs can be defined based on the coefficients of the linear combination. Considering that the value of individual graph edges and the value of probability T-A matrix are all non-negative, it is natural to consider the non-negative matrix factorization (NMF) (D. D. Lee & Seung, 1999) when identifying the bases of M. Two more reasons to use NMF: (1) NMF does not make the orthogonality assumption, thus it adapts to a wide range of matrix manifolds (Lu & Miao, 2016); (2) the bases obtained from NMF have physical meanings (Donoho & Stodden, 2003), called metagraph in the context of this study. NMF decomposes the total T-A matrix as $T_{n \times u} \approx W_{n \times k} \cdot H_{k \times u}$ . Note that the columns of W are actually the metagraphs and the axis of the metagraph space M. The columns of H are the corresponding coefficients of the linear combination of these metagraphs for the corresponding users, also can be seen as the corresponding coordinate point of the corresponding individual graph in the metagraph space M. The origin of the metagraph is an empty graph.

## 5. Study Case

### 5.1 Study area

In this paper, the Great Tokyo area was chosen as the study area (**Figure 5**). It includes Tokyo Metropolis and the surrounding area of prefecture Chiba, Ibaraki, Kanagawa, Saitama, Gunma, Yamanashi and Tochigi. Both metropolitan areas and rural areas are included in the study area.

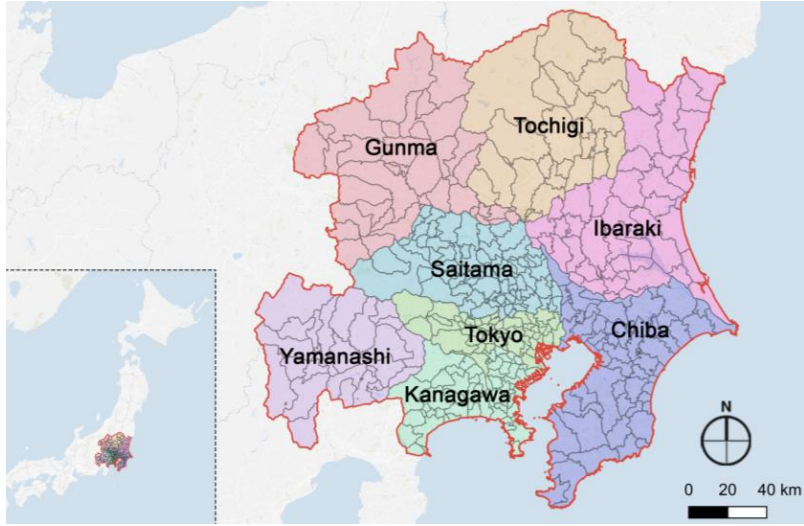

**Figure 5**. Study area

## 5.2 Dataset

In this study, we utilized GPS records from "Konzatsu-Tokei (R)" from 2013 January 1st to 2013 July 31st as the dataset. "Konzatsu-Tokei (R)" is an individual location dataset collected from mobile phones with enabled AUTO-GPS function under users' consent, through the "Docomo Map Navi" service provided by NTT DOCOMO, INC. It does not include specific individual information such as age or gender. The dataset we utilized in this research contains 11.67 billion GPS records collected from about 2.35 million mobile-phone users throughout Japan. Original GPS location data (latitude, longitude) are sent in about every minimum period of 5 minutes.

## 6.Result and discussion

### 6.1 Significant places detection

In this study ,we set *eps* = 30m and *minpt* = 10 for DBSCAN method. These parameters are approximating to the uncertainty in GPS positioning. They are recommended in related references (Huang, Chen, & Pan, 2020; Kisilevich, Mansmann, & Keim, 2010; Ostrikov, Rokach, & Shapira, 2013; Savvas & Tselios, 2016; Tang, Liu, Wang, & Wang, 2015; Viswanath & Pinkesh, 2006) and proved to be effective in detecting the significant places of individuals. There are totally 74,693 users who can be clearly detected the home location. There are totally 74,693 home places, 17,033 other frequently visit nighttime spots, 50,125 workplaces, 63,183 other frequently visit daytime spots, and 266,895 other significant places detected for the raw GPS records. **Figure 6** shows the spatial distribution and individual frequency distribution of significant places.

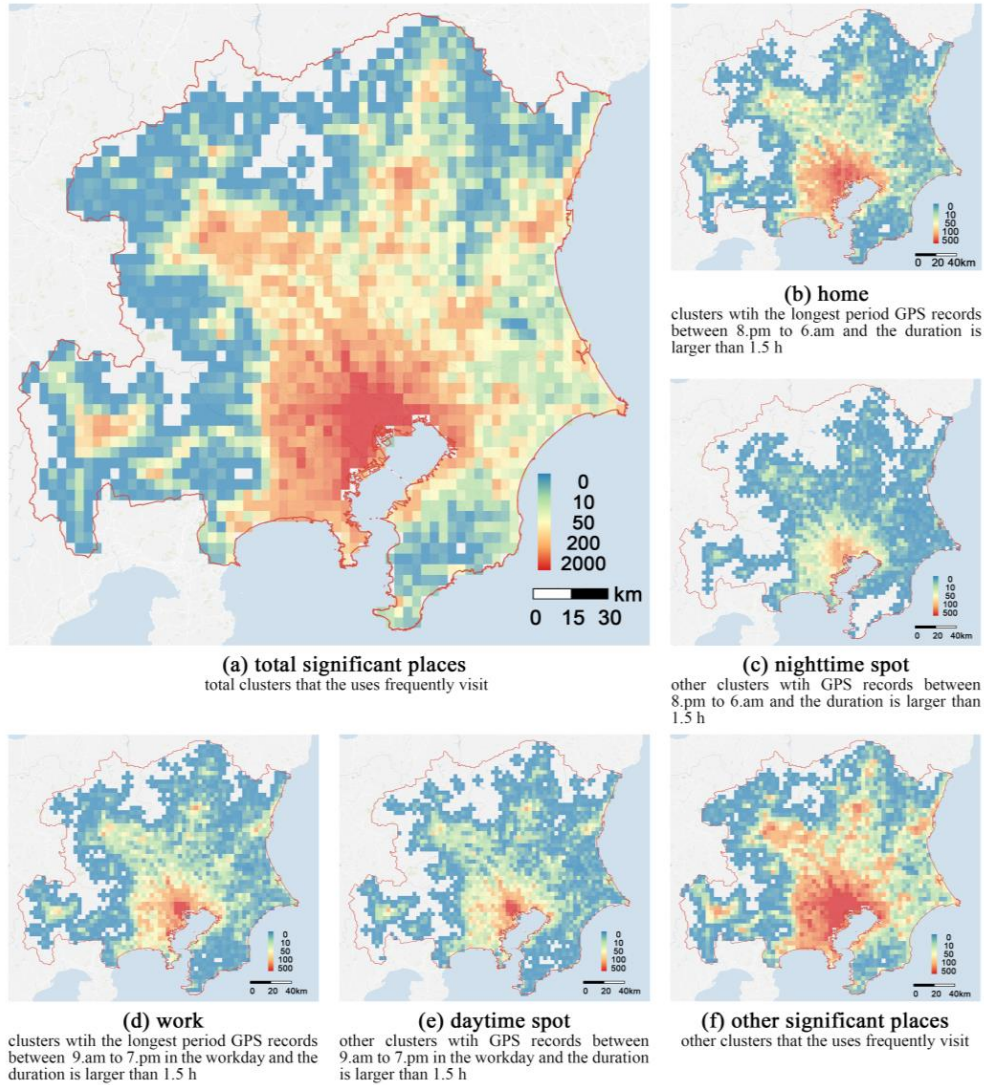

**Figure 6**. Heatmap and individual frequency distribution of significant places

To further validate our significant places estimation, we compared our results with the Mobaku census data in **Figure 7**. The Mobaku census data is provided by NTT DOCOMO, INC. It records the total population of grid units every 1 hour within Tokyo 23 wards from December $1^{st}$ (Sunday) to December $7^{th}$ (Saturday). The grid size is 500 m ×500 m. **Figure 7(a)** shows the average population of each grid from 03:00 to 05:00 within one week, which can be seen as the residential population of each grid. **Figure 7(b)** shows the number of home detected from GPS data. **Figure 7(c)** shows the average population of each grid from 09:00 to 17:00 within weekdays, which can be seen as the working population of each grid. **Figure 7(d)** shows the number of workplaces detected from GPS data. All the value has been normalized to be between 0 and 1. We can see that the detected population based on our home/work location estimation of significant places identification was comparable with the population information obtained from the census data.

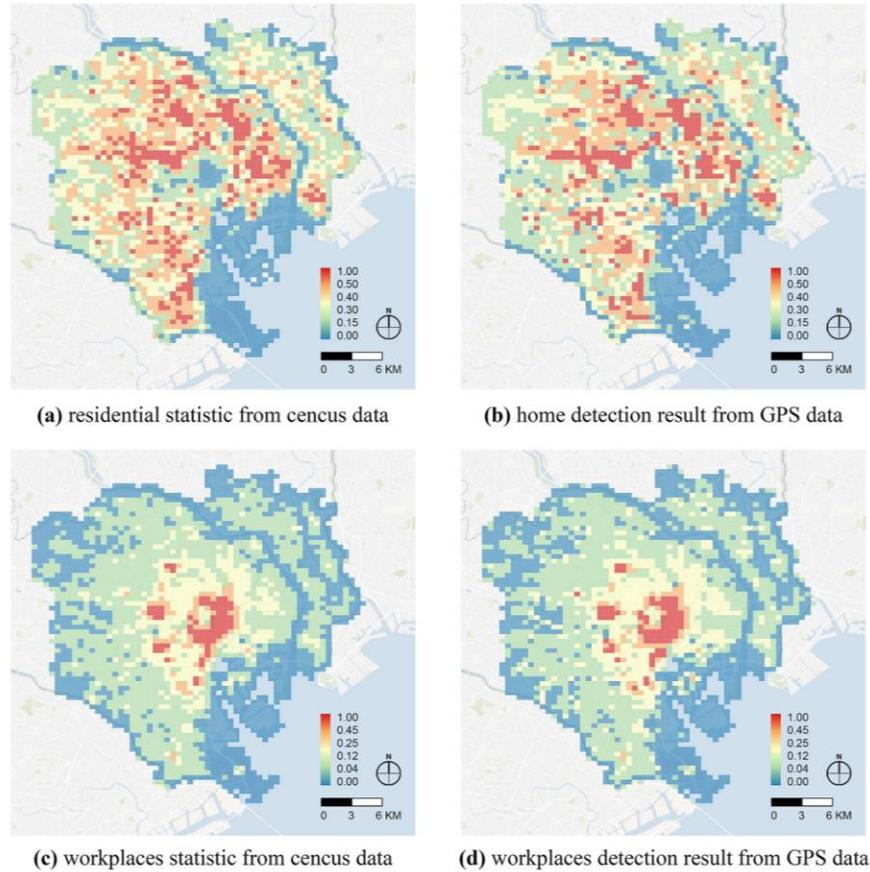

**Figure 7**. Comparison between estimated home/work population from GPS data and the actual population obtained from the census data

## 6.2 T-A matrix generation and NMF

Based on the generation principle we introduced above, we generated the support graph which describes the life pattern of 74,693 users within the research area from Jan $1^{st}$ 2013 to Jul $31^{st}$ 2013. There are 24 layers of the support graph and there are totally 3,871 edges of the support graph. 74,693 individual probability T-A matrices are generated based on the support graph. By merging the individual probability T-A matrix, we obtained a 3871×74639 total T-A matrix. After applying NMF methods, the total T-A matric is factorized into two matrices W (3871×3 ) and H (3×74639). Each column of W refers to one base of the metagraph space. Each column of H refers to one coordinate point (x,y,z) of one user in the metagraph space. **Figure 8(a)** shows the result of NMF. **Figure 8(b)** is the visualization of the significant pattern which the bases of the metagraph space represent. As each row of W refers to one edge of the support graph, that is, each row of W has its physical meaning, the larger value in W shows that the corresponding edge and the possibility of the corresponding activity are more significant. Here, we visualized the edges with values larger than 1. We can see that the bases of metagraph space represent three strong patterns: stay at home all day, stay at other places all day and stay at workplaces all day.

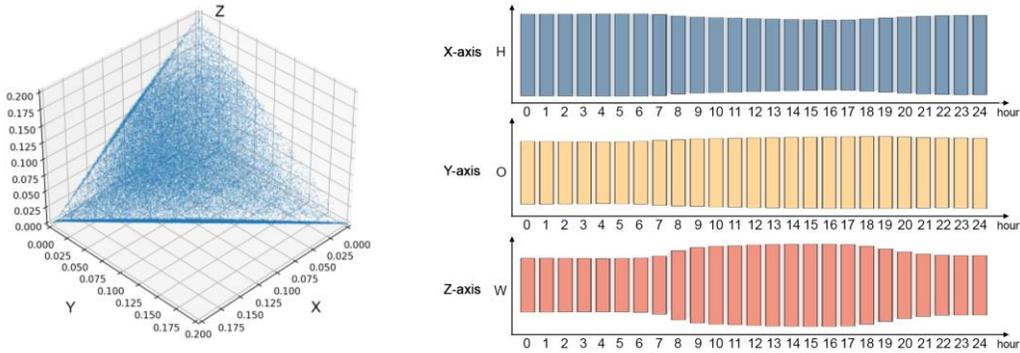

(a) The individual points in the metagraph space    (b) The strong pattern that the X,Y,Z axises of the metagraph space represent

**Figure 8**. MNF result

To validate the performance of our methodology, we compared the clustering result which directly applying K-means methods to 74,639 individual probability T-A matrix. In the directly K-Means method, we directly clustered the individual probability T-A matrices, while in NMF clustering, all the individual probability T-A matrices would be reduced the dimension into the metagraph space first. Here, we classified them into 3 groups by K-methods. **Figure 9** shows the comparison result. We can see that the proposed method can achieve a similar clustering result as the directly K-Means method. The 3 clustered groups have significant differences in the distribution of average cumulative time of staying at H, N,W,D and O. The 3 clustered groups also have significant differences in average probabilities of staying at home, workplaces and other places. Under the same computation situation, the directly K-Means cost 18,684 seconds for clustering 74,639 users into 3 groups, while the proposed methods only cost 486 seconds. The computation time of the proposed method is nearly 40th times faster than the directly K-Means method. The proposed method shows advantages in computation efficiency. Besides, when the number of clusters increases, it is almost impossible for the directly K-Means method to cluster the massive of users, as the computation tasks are too heavy. However, the proposed method can easily handle the clustering tasks of massive groups with as many users.

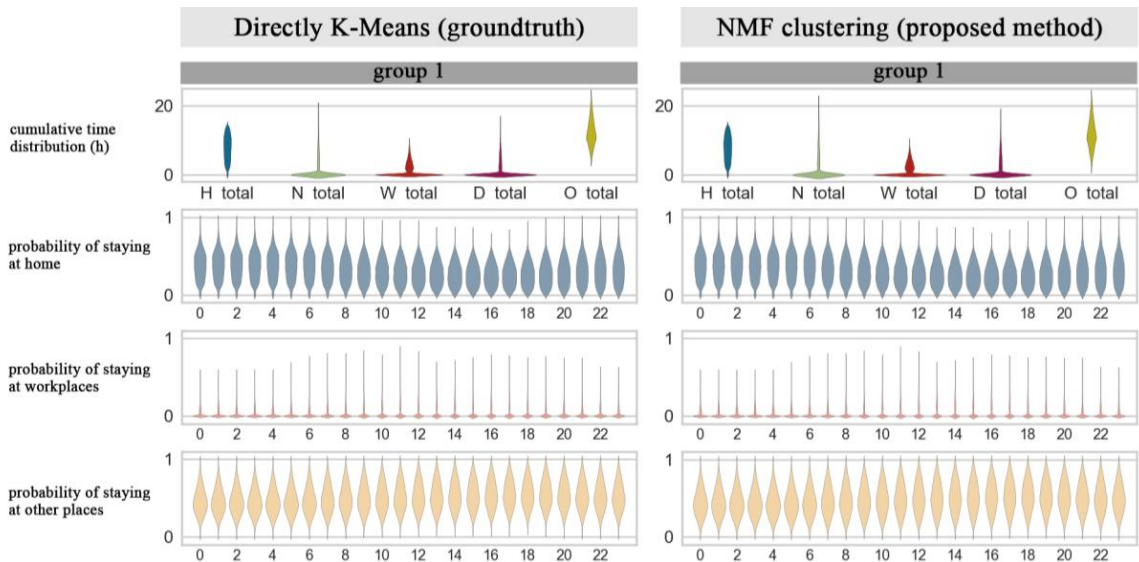

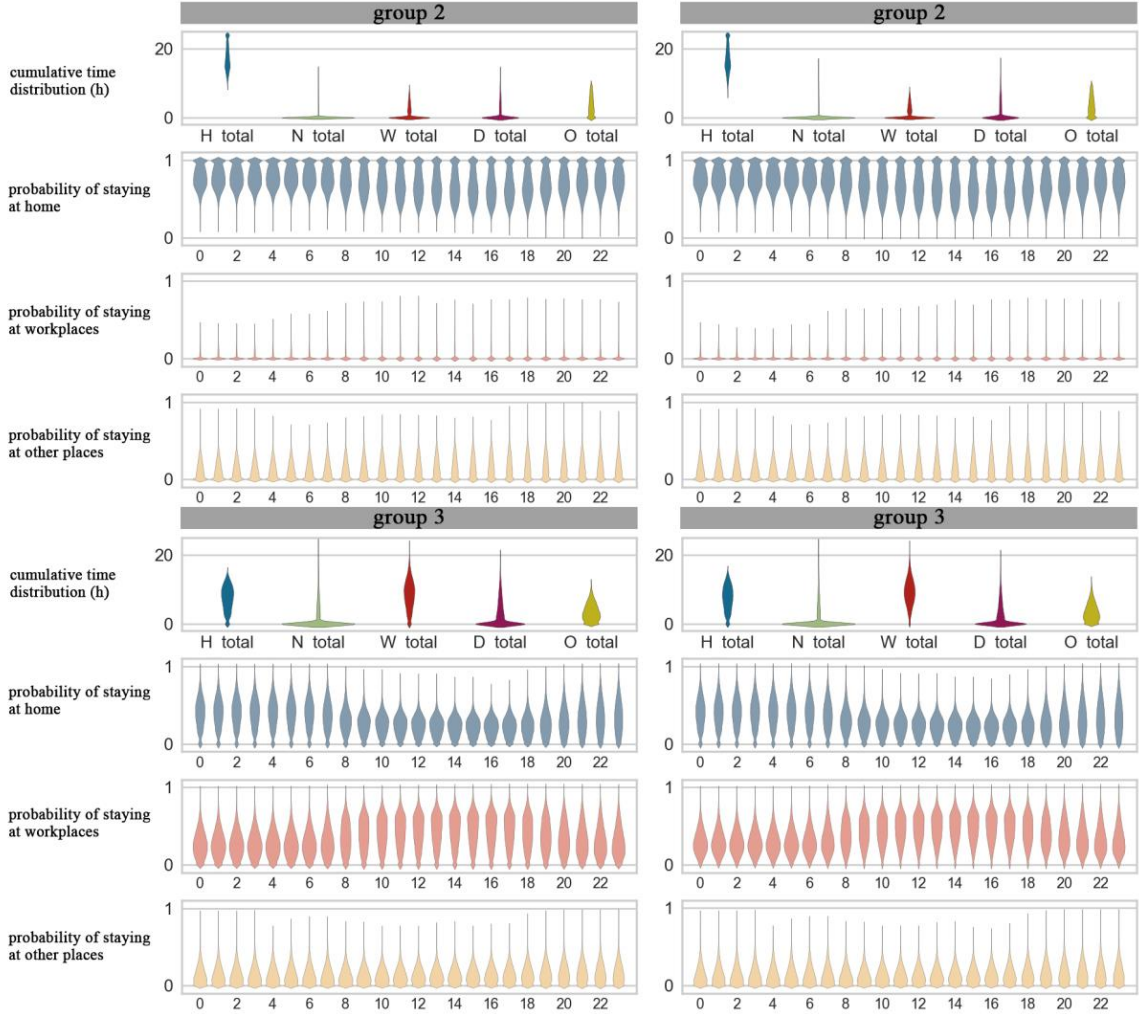

**Figure 9**. Comparison of directly KMeans and MNF methods

## 6.3 Group characteristics of life pattern

In this section, we explore the life pattern similarity in the metagraph space. As the high-dimension life pattern data are transformed to one point in the metagraph space, Euclidean distance can be introduced to measure the similarities between the points of this metagraph space. Traditional clustering methods such as K-means can also be incorporated into the metagraph space. Here, the distortion elbow method is used for determining the optimal value of $k$ in K-Means which is introduced in the metagraph space. It is calculated as the average of the squared distances from the cluster centers of the respective clusters. As 7 is the point of inflection on the curve (**Figure 10**), we classify all the users into 7 groups.

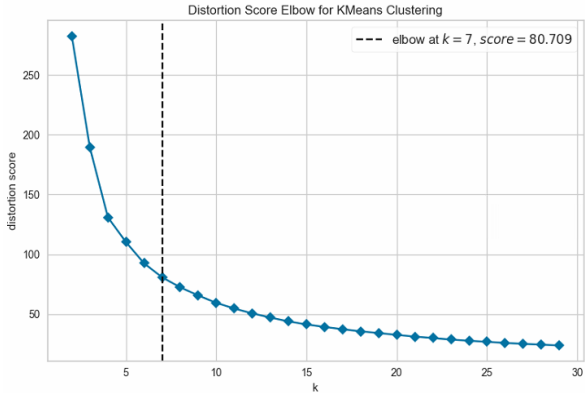 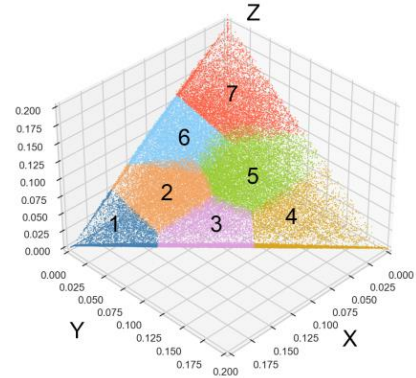

(a) Elbow method result  (b) 7 clusters by K-means in the metagraph space

**Figure 10**. Determine the number of cluster in Kmeans by elbow methods

**Table 2** shows the life pattern characteristics of the central point of each group and gives a general description of the life pattern of each group. The X-axis refers to 24 hours of a day. Each block refers to one node in the support graph, that is, one significant place of each hour. The height of the block in the Y-axis direction refers to the probability of which the users stay in the corresponding significant place. Here, we only visualized the block whose probability is larger than 1%. We can see that the users who have larger X values in the metagraph space tend to have a higher probability to stay at home. The users who have a larger Y value in the metagraph space tend to have a higher probability to stay at the other significant places. The users who have a larger Z value in the metagraph space tend to have a higher probability to stay at the workplace. From **Table 2** we can see that our proposed method can effectively cluster people with similar life patterns in which significant places semantics, time sequential properties and frequency are considered. Besides, the measurement of life pattern similarity becomes possible, for it is concerted into the Euclidean distance of the metagraph space. The users whose corresponding points in the metagraph space are closer tend to have a more similar life pattern.

**Table 2**. Life pattern characteristics of each group

| group characteristics | description |
|---|---|
| (chart showing mostly H (home) stay across 24 hours with minor O activity) | **Group 1. Home stayer: 85% Stay-at-home**<br><br>The users of this group spent most of their time at home. They almost have no workplace and seldom go to other places. This group may refer to the old people or housewives who stay at home all day. |

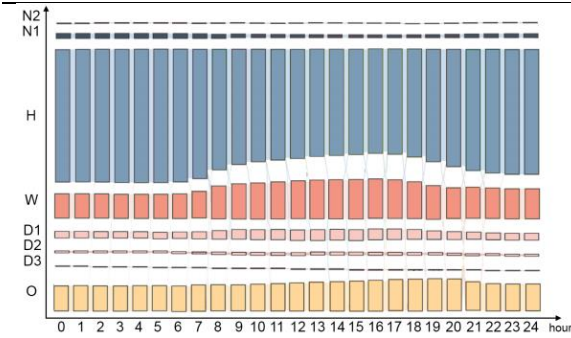

**Group 2. Home stayer: 60%-Stay-at-home**

The users of this group have clear workplaces but they don't go to the workplaces often. This group may refer to the people who telework at home and go to the company sometimes.

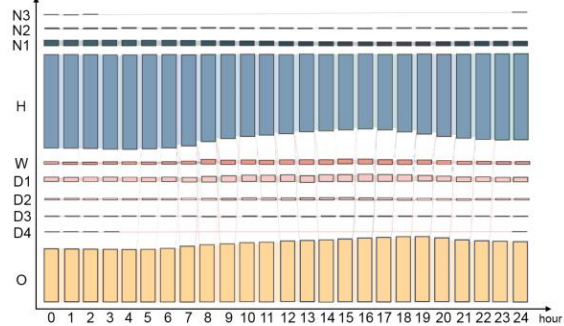

**Group 3. Home stayer: 50%-stay-at-home and 40%-stay at-other**

The users of this group spend most of their time in the home or other places. They almost have no workplace.

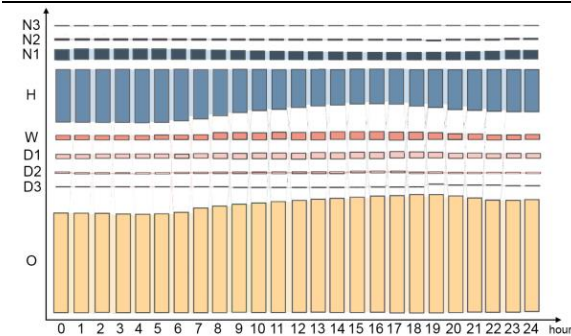

**Group 4. Traveler: 70%-stay-at-other**

The users of this group spend most of their time in multiple other significant places.

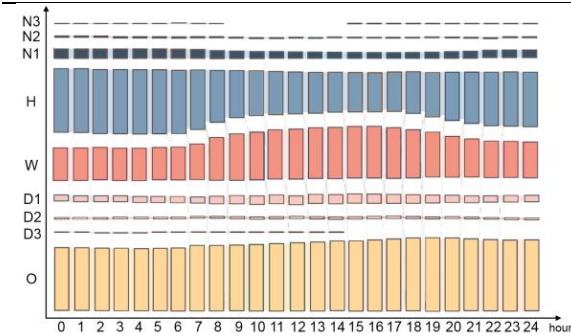

**Group 5. Work-life balance**

The users of this group spend roughly equal time at home, work and other places. Their working times are relatively short and they have much leisure time to go to other places.

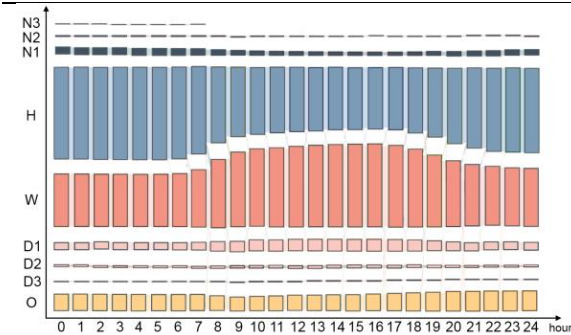

**Group 6. Regular office workers**

The users of this group usually stay at home at nighttime and go to work in the daytime. They are typical office workers and have regular schedules. Little leisure time is spent in other places.

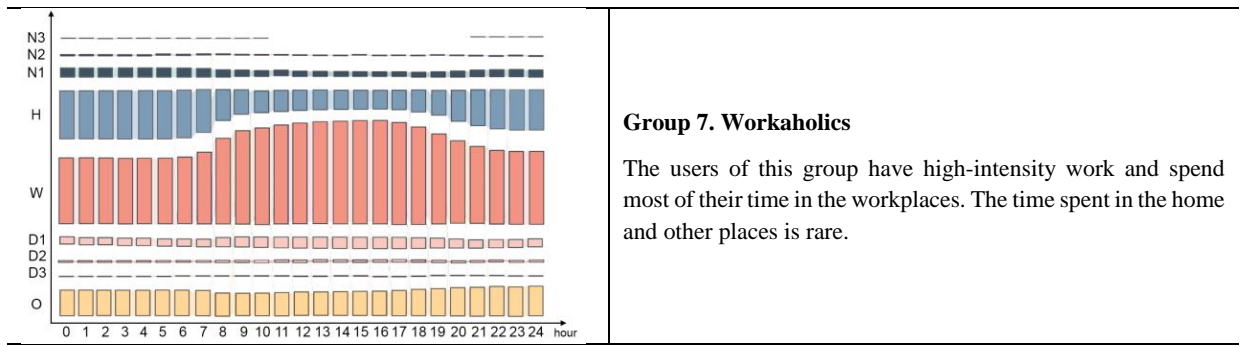

**Group 7. Workaholics**

The users of this group have high-intensity work and spend most of their time in the workplaces. The time spent in the home and other places is rare.

## 6.4 Regional characteristics of life pattern

To explore the spatial distribution characteristics of different life patterns, we visualized the standard deviation of the population percentage of 7 groups in the study area (**Figure 11**). To avoid bias, the grids whose total population is fewer than 5 are excluded. We can see that the standard deviation of the population percentage of the 7 groups in metropolitan areas is small while the standard deviation in the outlying area is high. This indicates that the types of life pattern in the metropolitan area are more diverse and the one in the outlying area is relatively single. This could because the demographic composition in metropolitan areas is more even and the demographic composition in the outlying area is more imbalanced, so the standard deviation of population percentage of life pattern could be higher.

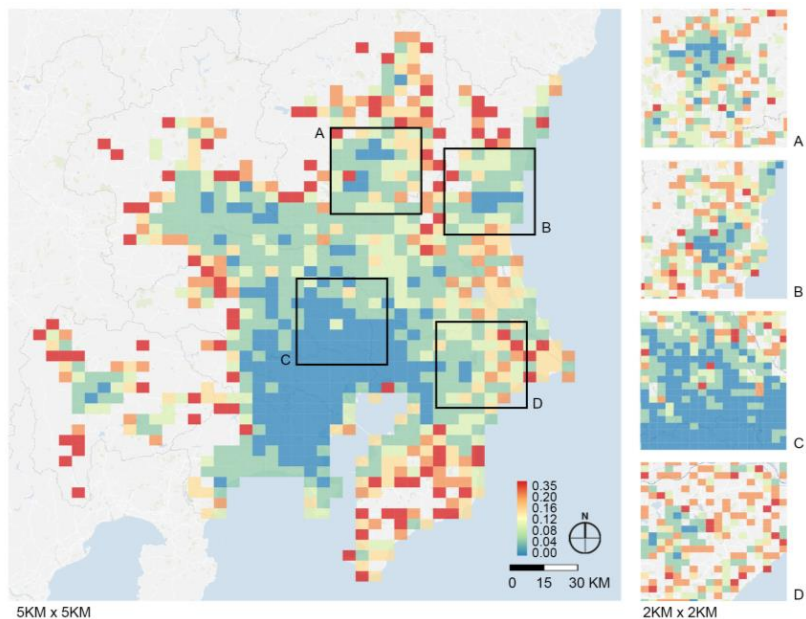

**Figure 11.** The standard deviation of population percentage of 7 groups

We further compared the correlation between the population percentage of the 7 groups of grids. **Table 3** shows the statistical significance and the Pearson correlation coefficient between the population percentage of the 7 groups. There are significant negative correlations between most groups. This is in line with the common perception that when the proportion of a certain group is high, the proportion of the other groups will decrease. There are no significant correlations between group 1 and group 2, group 4 and group 5, group 4 and group 7, group 5 and group7. As group 1 trend to stay at home most of time, while group 4 trend to stay at other significant places at most of time and group 7 trend to stay at workplaces at most of time, we can infer that there are some degree of correlations between the possibility of people staying at home and the possibility of people going to work, and there is some degree of correlation between the possibility of people staying at home and the possibility of people going to other significant places such as shopping mall, gym, leisure and entertainment. However, there is no strong correlation between the possibility of people going to work and the possibility of people going to other significant places such as shopping malls, gyms, leisure and entertainment.

**Table 3**. the statistical significance and the Pearson correlation coefficient between the population percentage of the 7 groups

|  |  | Group 1 | Group 2 | Group 3 | Group 4 | Group 5 | Group 6 | Group 7 |
|---|---|---|---|---|---|---|---|---|
| **Group 1** | Correlation Coefficient | 1 | **-.052** | -.103** | -.285** | -.293** | -.118** | -.195** |
|  | Sig.(2-tailed) |  | **.105** | .001 | .000 | .000 | .000 | .000 |
| **Group 2** | Correlation Coefficient | **-.052** | 1 | -.137** | -.247** | -.263** | -.122** | -.227** |
|  | Sig.(2-tailed) | **.105** |  | .000 | .000 | .000 | .000 | .000 |
| **Group 3** | Correlation Coefficient | -.103** | -.137** | 1 | -.180** | -.306** | -.123** | -.194** |
|  | Sig.(2-tailed) | .001 | .000 |  | .000 | .000 | .000 | .000 |
| **Group 4** | Correlation Coefficient | -.285** | -.247** | -.180** | 1 | **.011** | -.187** | **-.036** |
|  | Sig.(2-tailed) | .000 | .000 | .000 |  | **.723** | .000 | **.261** |
| **Group 5** | Correlation Coefficient | -.293** | -.263** | -.306** | **.011** | 1 | -.218** | **.020** |
|  | Sig.(2-tailed) | .000 | .000 | .000 | **.723** |  | .000 | **.539** |
| **Group 6** | Correlation Coefficient | -.118** | -.122** | -.123** | -.187** | -.218** | 1 | -.230** |
|  | Sig.(2-tailed) | .000 | .000 | .000 | .000 | .000 |  | .000 |
| **Group 7** | Correlation Coefficient | -.195** | -.227** | -.194** | **-.036** | **.020** | -.230** | 1 |
|  | Sig.(2-tailed) | .000 | .000 | .000 | **.261** | **.539** | .000 |  |

## 6.5 Temporal changes of life pattern

We separated the individual probability T-A matrix to weekdays' probability T-A matrix and weekends' probability T-A matrix of each user based on the support graph generated in **section 6.2**. 74689 users have both weekdays' probability T-A matrix and weekends' probability T-A matrix. Based on the same method introduced above, we obtain a 3871×149378 total T-A matrix. After applying NMF methods, the total T-A matrix is factorized into matrix W (3871×3) and matrix H (3×149378). **Figure 12** shows the result. We can see that although the total T-A matrix and the basis of metagraph space are different from 6.2, the base of metagraph space shows a similar pattern as 6.2. This means that if the factorized matrix changes, the basis of the metagraph space will change but its physical properties will remain the same. The proposed method shows robustness.

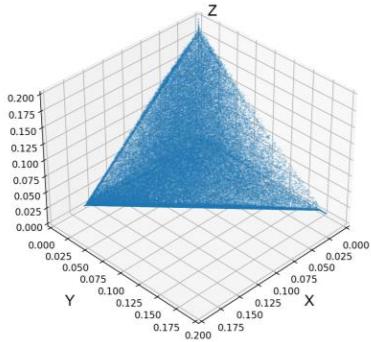 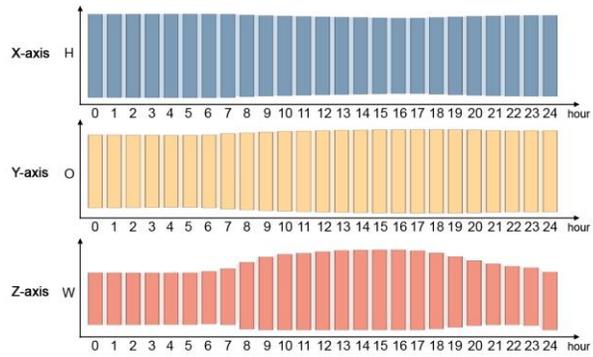

**(a)** The individual point in the metagraph space   **(b)** The strong pattern that the X,Y,Z axises of the metagraph space represent

**Figure 12.** MNF result of weekday and weekend average probability matrix

After applying the same K-Means classifier as **section 6.3**, we classified people into 7 groups as **Table2**. We conducted statistics on how people change their life patterns groups on weekdays and weekends (**Figure 13**). Group1 and Group 2 (blue bar in **Figure 13**) tend to stay at home more frequently; Group3 and Group 4 (yellow bar in **Figure 13**) tend to stay at other significant places more frequently; Group 5 (green bar in **Figure 13**) have balances between home, work and other significant places; Group 6 and group 7 trend to stay at workplaces more frequently. From the aspect of people who stay at workplaces frequently on weekdays, from weekdays to weekends, 43.48% of group 6 change to group 1, 20.2% of group 6 change to group 2. In contrast to this, 7.00% of group 6 change to group 3, 0.93% of group 6 change to group 4. Similarly, from weekdays to weekends, 21.38% of group7 shift to group 1, 12.32% group 7 shift to group 2, while 12.9% of group 7 shift to group 3, 11.64% of group 7 shift to group 4. For people who stay at workplaces frequently on weekdays, the people who transform to stay at home on weekends are more than those who transform to go to other places on weekends. From the aspect of people who stay at other significant places frequently on weekdays, from weekdays to weekends, 24.62% of group 3 change to group 1, 2.70% of group 3 to group 2. In contrast, 0.05% of group 3 change to group 6, 0.03% of group 3 change to group7. Similarly, 7.32% of group 4 change to group1, 0.62% of group 4 change to group 2, while 0.01% of group 4 change to group 6, 0.069% of group 4 change to group 7. For people who stay at other significant places frequently on weekdays, the people who transform to stay at home on weekends are also more than those who transform to go to workplaces on weekends.

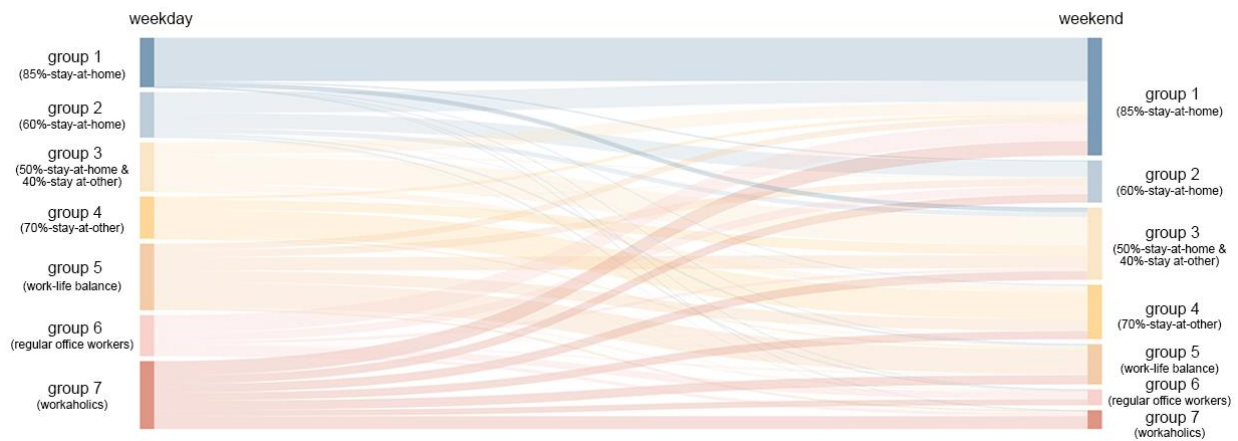

**Figure 13.** transformation of life patterns groups between weekdays and weekends

# 7. Conclusion

In this research, based on millions of GPS records, we proposed a novel life pattern clustering method considering spatial-temporal similarity integrating significant places semantics, time sequential properties and frequency. Against the challenges in life pattern clustering with regard to diversity, uncertainty and high-dimension, we introduced a metagraph-based matrix to present the diverse and uncertain life patterns of the individual in a unity format. The Non-negative-factorization-based method was utilized for reducing the dimension. Based on the clustering result, we further analyzed the group characteristics of human life patterns during different periods and different regions. The result shows that our proposed method can efficiently identify the groups have similar life pattern and takes advantages in computation efficiency and robustness. The types of life pattern in the metropolitan area are more diverse and the one in the outlying area is relatively single.

Our proposed method has positive significance in urban planning. First, it provides an effective measurement for life pattern similarity. It is convenient for urban planners to directly compare the differences and diversification of life patterns in region scale, which helps in policy making for urban problems such as traffic congestion, home-work separation, public facilities mismatch and regional unbalanced. Secondly, it reveals the representative life pattern groups. The representative activities and travel demand of these groups need to be considered in infrastructure planning and services improvement to provide them with better urban systematic support. Thirdly, as the complex life patterns including significant places semantics, time sequential properties and frequency are integrated into one vector in the metagraph space, it provides a simplified indicator the urban modeling and quantitative evaluation.

There are still some limitations. Despite the dataset we used contains a huge amount of GPS records of massive users, it only accounts for one percent of the total population of Japan. This could bring some degree of bias in population estimation. In the future, a detailed survey is required on individual life patterns. Such demographic information will contribute to further population estimation and quantitative analysis.

# Declaration of Interest

None

# Contribution

Haoran Zhang: Conceptualization, Methodology, Formal analysis; Wenjing Li: Writing, Algorithm development, Conceptualization, Formal analysis; Peiran Li, Yuhao Yao, Mariko Shibasaki, Xuan Song, Ryosuke Shibasaki: Supervision